# CONTRIBUTIONS OF GEORGE J. PAPADOPOULOS

J. T. Devreese[(a),(b)]

[(a)]TFVS, Departement Fysica, Universiteit Antwerpen, B-2610 Antwerpen, Belgium
[(b)]eiTT/COBRA Inter-University Research Institute, Physics Department, Eindhoven University of Technology, P.O. Box 513, 5600 MB Eindhoven, The Netherlands
E-mail: jozef.devreese@ua.ac.be

**Abstract**

I review some of the scientific work of George Papadopoulos in the context of the Greek cultural tradition and modern theoretical physics. The main emphasis is on his works on path integrals and their applications. The review is closed by an excursus on the polaron physics, where the path-integral approach has been proven to be a method of excellence.

## 1. Preface

Traditions of theoretical studies in Greece (see Fig. 1) are related to such names as Thales, Pythagoras, Plato, Aristotle, Archimedes… Those traditions have been continued in recent time by George Papadopoulos. I review some of the scientific work of George Papadopoulos, with whom I had the pleasure to collaborate over longer time periods, including at Antwerpen. On this occasion, I wish George many new creative achievements in the future years.

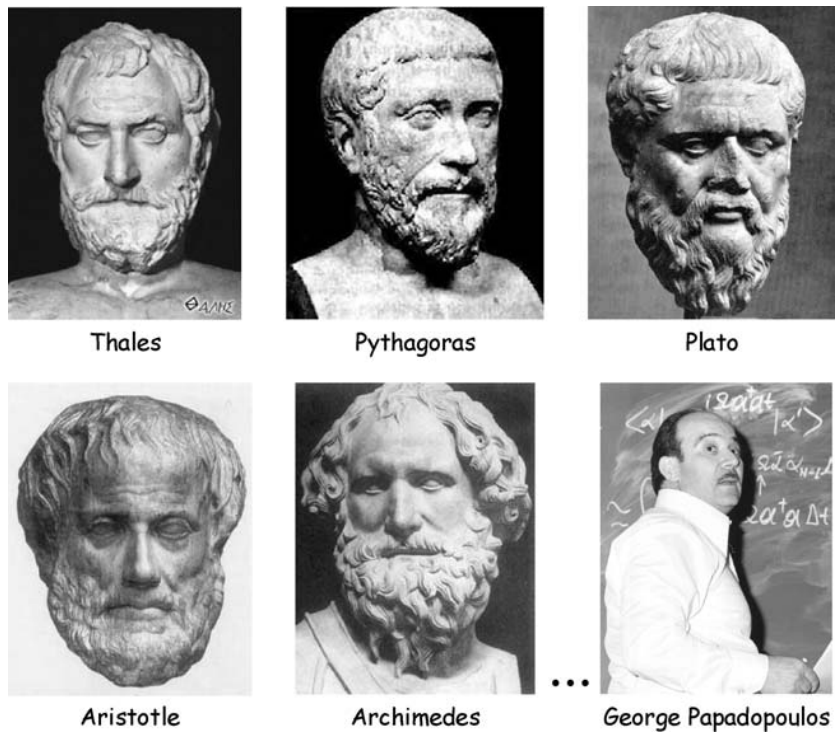

Fig. 1. Traditions of theoretical studies in Greece.

## 2. Path integrals and their applications

Path integrals and their applications in quantum, statistical and solid state physics have been the subject of the book [1], co-authored by George Papadopoulos. *A great number of paths of the Labyrinth[2] (Fig. 2) can be compared to an infinite number of paths for a path integral.*

Among the original contributions of George his excursions on quantum dynamics and path integrals must be mentioned.

The path integral for the partition function, which determines the free energy, is calculated over closed paths. *An example of a closed path is the trajectory of the Argonauts' journey[2] (Fig. 3).*

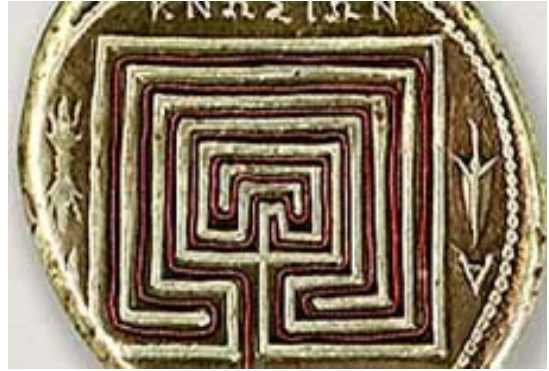

Fig. 2. "…Now the Labyrinth which Daedalus constructed was a chamber that with its tangled windings perplexed the outward way." (*Apollodorus, Library and Epitome, 3.1.4*)
(From
http://www.eichfelder.de/kulte/labyrint/vortrag.html)

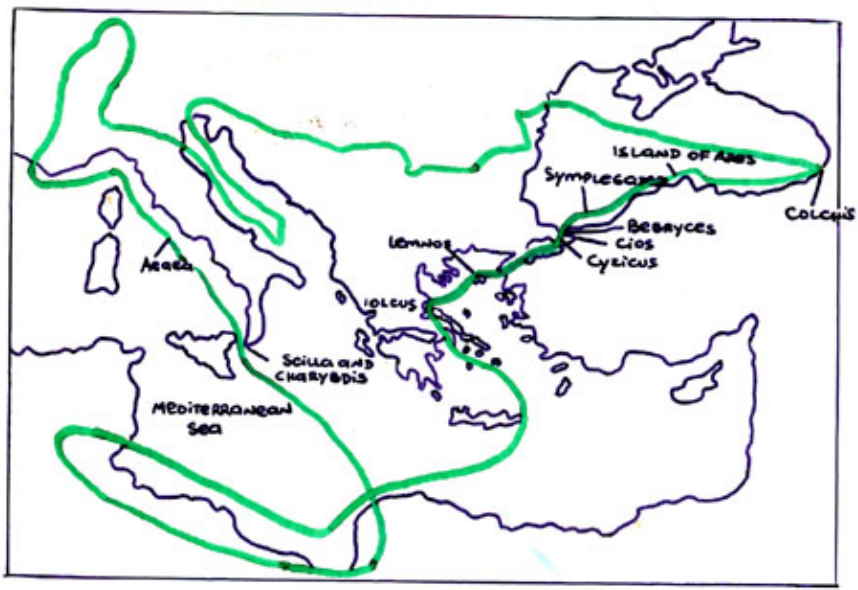

Fig. 3. "Beginning with thee, O Phoebus, I will recount the famous deeds of men of old, who, at the behest of King Pelias, down through the mouth of Pontus and between the Cyanean rocks, sped well-benched Argo in quest of the golden fleece."
(*Apollonius Rhodius, Argonautica*) (Map from http://lukio.vimpeli.fi/italia/argoviag.htm).

He treated quantum dynamics of dissipative Lagrangians, proposed a systematic approach for the derivation of the path-integral propagator for a constrained particle described by a dissipative Lagrangian [3]. *The role of dissipation is demonstrated in the story of Pelops and Oenomaus (Fig. 4)*. As friction heated the axle of Oenomaus' chariot, the wax linchpins melted, and when the wheels eventually fell off, the whole thing broke apart.

Further, George developed a path integral description of an electron gas in a random potential [4] and analyzed functional integrals with applications in polymer physics and for spin-Bose systems.

Feynman's path-integral method had already been extremely successful in tackling nonrelativistic quantum-mechanical problems. Although, in principle, there was nothing to prevent the application of the method to obtain a formal expression for the propagator of the Dirac equation, in practice a variety of mathematical difficulties had prevented direct actual calculations. In 1976 George co-authored a paper [5] in which the propagator for a free fermion and for a fermion in a magnetic field was directly derived in the path-integral formalism. The propagator for a Dirac electron in a constant magnetic field was indirectly obtained by evaluating a world-line (space-time path) integral. The corresponding spectrum was then extracted from an auxiliary propagator.

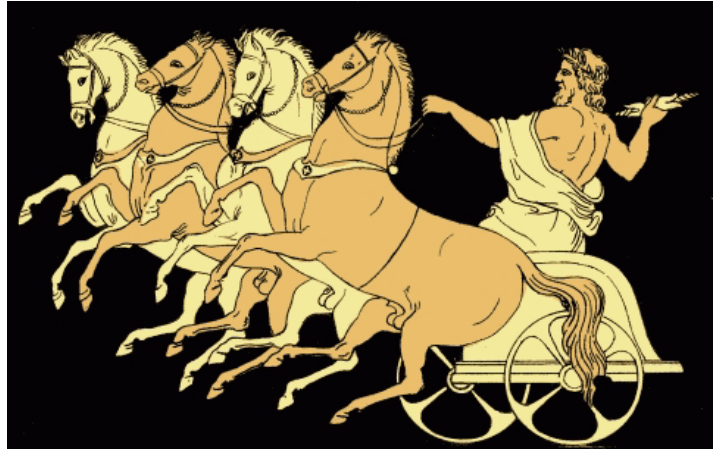

Fig. 4. "And therein were fashioned two chariots, racing, and the one in front Pelops was guiding, as he shook the reins, and with him was Hippodameia at his side, and in pursuit Myrtilus urged his steeds, and with him Oenomaus had grasped his couched spear, but fell as the axle swerved and broke in the nave, while he was eager to pierce the back of Pelops." (*Apollonius Rhodius, Argonautica*) (The Chariot of Zeus, from *Stories from the Greek Tragedians* by Alfred Church, 1879. http://en.wikipedia.org/wiki/Image:The_Chariot_of_Zeus_-_Project_Gutenberg_eText_14994.png)

### 3. Spectrum of Scientific Activities

Other works of George demonstrate a broad spectrum of his fruitful scientific activities.

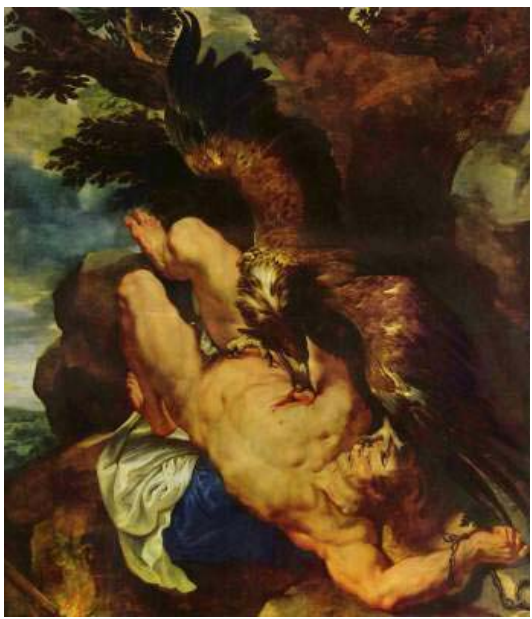

Problems of relativistic physics remained one of George's key interests over his scientific career. He and his co-authors provided a useful approximation to treat the Dirac equation with a scalar potential and a fourth component of the vector potential of rectangular shape [6] and demonstrated the analytic advantages of those potentials [7].

Fig. 5. "Thou firmament of God,
and swift-winged winds,
Ye springs of rivers, and of ocean waves
That smile innumerous! Mother of us all,
O Earth, and Sun's all-seeing eye, behold,
I pray, what I, a God, from Gods endure."
(*Aeschylus, Prometheus Bound*,
http://www.bartleby.com/8/4/1.html).
(Picture by P. P. Rubens, from
http://www.artprints-on-demand.co.uk/
noframes/rubens/prometheus.htm)

*Prometheus Bound provides an example of a confinement (Fig. 5).*

Several of the works of George and his co-workers deal with the interaction of light with matter. He analyzed the interaction of radiation with atoms [8], light-scattering properties of linear polymers, energy exchange between parametric modes in a nonlinear optical medium, the energetics of a system consisting of radiation and a two-level atom in an ideal resonant cavity [9], the amplitude and phase of the acoustic effect.

Another field of his research interests is in the physics of tunneling, where he provided important ingredients for the understanding of time-dependent quantum tunneling via crossover processes [10] and of photon induced tunneling oscillations in a double quantum well [11].

George and his co-authors also extensively applied fundamental theoretical approaches to problems of materials physics. For example, they characterized polycrystalline polyamine copper dinitrate complexes [12] and iron oxide pigments: hematite, goethite and magnetite [13] using photoacoustic, EPR and electrical conductivity investigations.

## 4. Polarons

The polaron concept is of interest, not only because it describes the particular physical properties of an electron in polar crystals and ionic semiconductors, but also because it is an interesting field-theoretical model consisting of a fermion interacting with a scalar boson field.

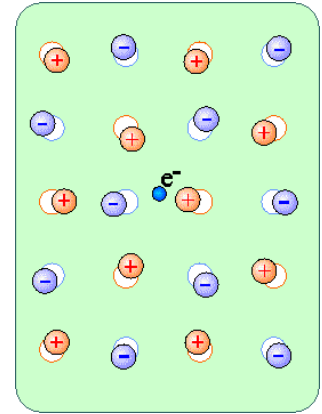

A conduction electron (or hole) together with its self-induced polarization in a polar crystal forms a quasiparticle, which is called a polaron [14-17]. Properties of polarons have attracted increasing attention due to their relevance to physics of conjugated polymers, colossal magnetoresistance perovskites, high-$T_c$ superconductors, layered $MgB_2$ superconductors, fullerenes, quasi-1D conductors, semiconductor nanostructures.

Fig. 6. An artist's view of a polaron. (From Ref. [18])

A conduction electron repels the negative ions and attracts the positive ions (Fig. 6). A self-induced potential arises, which acts back on the electron and modifies its physical properties. The polaron coupling constant was introduced by Fröhlich [19]:

$$\alpha = \frac{e^2}{\hbar}\sqrt{\frac{m_b}{2\hbar\omega_{LO}}}\left(\frac{1}{\varepsilon_\infty} - \frac{1}{\varepsilon_0}\right),$$

where $\omega_{LO}$ is the long-wavelength frequency of a longitudinal optical (LO) phonon; $\varepsilon_\infty$ and $\varepsilon_0$ are, respectively, the electronic and the static dielectric constant of the polar crystal, $m_b$ is the electron (hole) band mass.

Feynman's all-coupling path-integral treatment [20,21] is based on his suggestion to formulate the polaron problem in the Lagrangian form of quantum mechanics and then to eliminate the phonon field from the propagator. As a result, the polaron problem is formulated [20] as an equivalent one-particle problem in which the interaction, non-local in time or "retarded", is between the electron and itself:

$$\langle 0, \beta | 0, 0 \rangle = \int D\mathbf{r}(\tau) \exp\left[ -\frac{1}{2}\int_0^\beta \dot{\mathbf{r}}^2 d\tau + \frac{\alpha}{2^{3/2}} \int_0^\beta \int_0^\beta \frac{e^{-|\tau-\sigma|}}{|\mathbf{r}(\tau) - \mathbf{r}(\sigma)|} d\tau d\sigma \right]$$

with $\beta = 1/(k_B T)$.

Feynman introduced a variational principle for path integrals to study the polaron. He simulated the interaction between the electron and the polarization modes by a harmonic interaction between a hypothetical particle and the electron (Fig. 7).

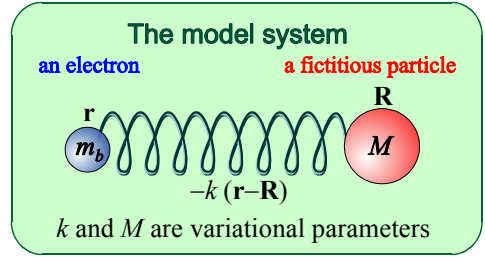

Fig. 7. The Feynman model system for a variational study of the polaron.

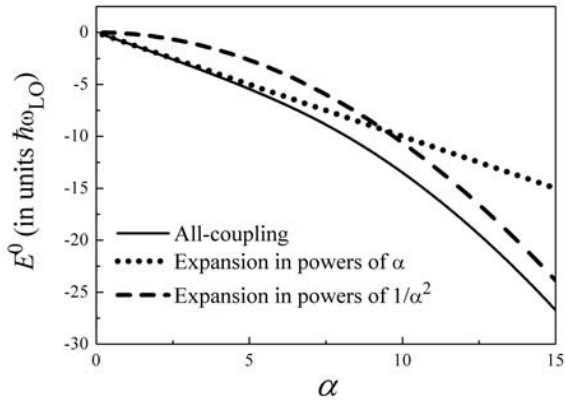

Fig. 8. Feynman-polaron energy as a function of $\alpha$: the all-coupling theory.

Applying the variational principle for path integrals resulted in an upperbound for the polaron self-energy at all $\alpha$, which at weak and strong coupling gave quite accurate limits (see Fig. 8). Feynman obtained the smooth interpolation between weak and strong coupling for the ground-state energy. Over the years, the Feynman polaron model remained in many respects the most successful approach to this problem.

At zero temperature and in the weak-coupling limit, the optical absorption is due to the elementary polaron scattering process, schematically shown in Fig. 9. An incoming photon is absorbed by a polaron. The polaron emits a phonon during the absorption process and takes recoil energy from the incident light.

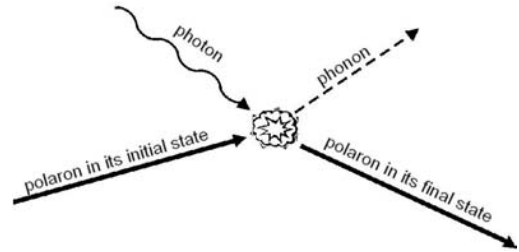

Fig. 9. Elementary polaron scattering process. (From Ref. [22]. © 2003 by the American Institute of Physics)

In the weak-coupling limit ($\alpha \ll 1$) the polaron absorption coefficient at zero temperature can be expressed in terms of elementary functions in two limiting cases [23,24]:
at high densities

$$\Gamma(\omega) = \frac{1}{\epsilon_0 nc} \frac{2^{1/2} N^{2/3} \alpha}{(3\pi^2)^{1/3}} \frac{e^2}{(\hbar m_b \omega_{LO})^{1/2}} \frac{\omega - 1}{\omega^3} \Theta(\omega - 1)$$

and at low densities

$$\Gamma(\omega) = \frac{1}{\epsilon_0 nc} \frac{2Ne^2 \alpha}{3 m_b \omega_{LO}} \frac{(\omega - 1)^{1/2}}{\omega^3}$$

where $\omega$ is the frequency of the incident light represented in units of $\omega_{LO}$.

The optical absorption of a single large polarons at arbitrary coupling has been derived using the path-integral method [25, 26] (Fig. 10). At larger coupling, $\alpha \geq 5.9$, the polaron can undergo transitions toward a relatively stable internal excited state called the "relaxed excited state" (RES). The RES peak in the spectrum also has a phonon sideband, which is related to a Franck–Condon (FC)-type transition. The RES peak is very intense compared with the FC peak.

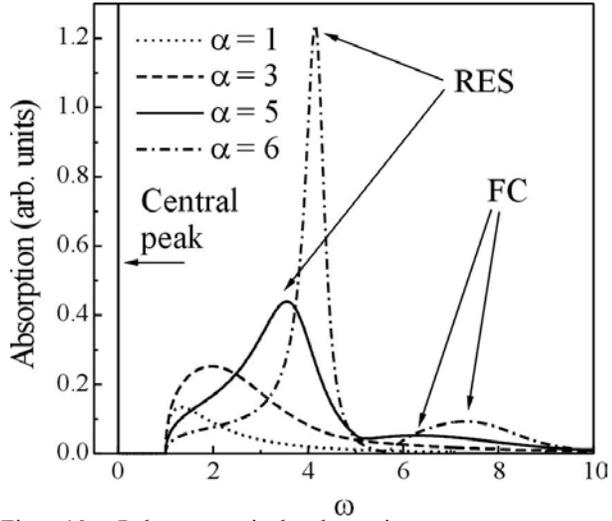

Fig. 10. Polaron optical absorption spectra at zero temperature, calculated within the path integral method [25, 26] for different values of $\alpha$. A $\delta$-like central peak is schematically shown by a vertical line. (From Ref. [22]. © 2003 by the American Institute of Physics)

Calculations of the optical conductivity for the Fröhlich polaron performed within the diagrammatic Quantum Monte Carlo method [27] (Fig. 11), fully confirm the results of the path-integral variational approach [25] at $\alpha \leq 3$. In the intermediate coupling regime $3 < \alpha < 6$, the low-energy behavior and the position of the maximum of the optical conductivity spectrum of Ref. [27] follow well the prediction of Ref. [25]. There are the following qualitative differences between the two approaches in the intermediate and strong coupling regime: in Ref. [27], the dominant peak broadens and the second peak does not develop, giving instead rise to a flat shoulder in the optical conductivity spectrum at $\alpha = 6$. This behavior can be attributed to the optical processes with participation of two or more phonons. The nature of the excited states of a polaron needs further study.

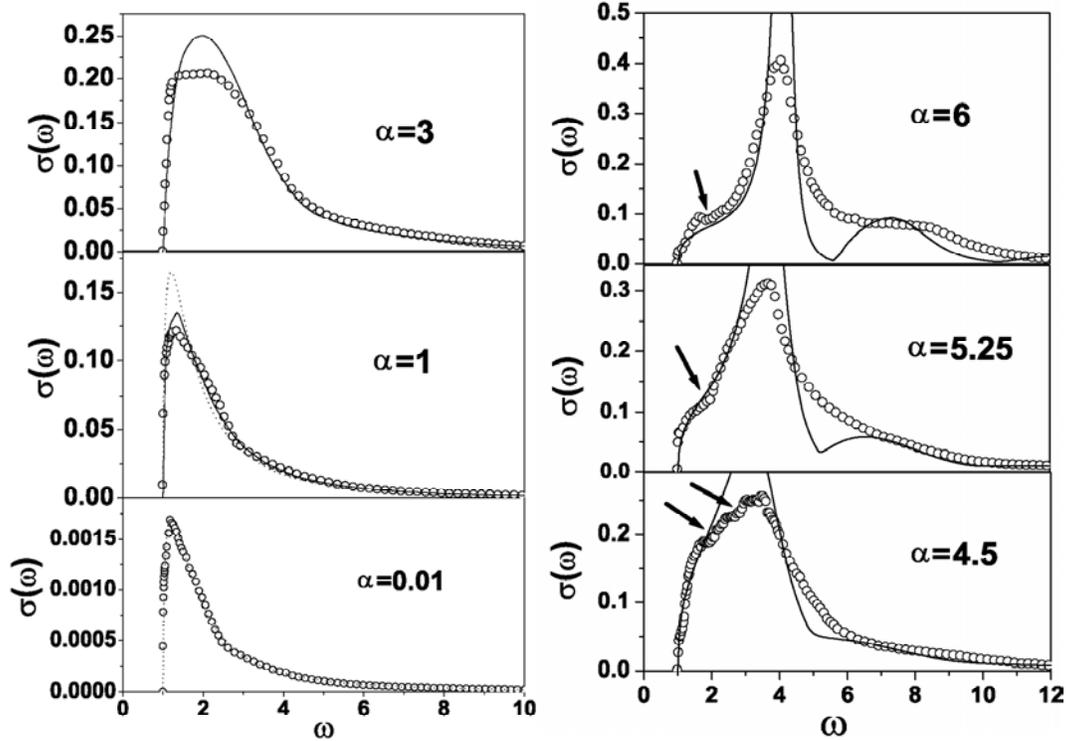

Fig. 11. Left-hand panel: Optical conductivity spectra for the weak-coupling regime (open circles) compared to the second-order perturbation theory (dotted lines) and arbitrary-coupling path-integral theory [25] (solid lines). Right-hand panel: Optical conductivity spectra for the intermediate coupling regime (open circles) compared to the arbitrary-coupling path-integral theory [25] (solid lines). Arrows point to the anomalies in absorption spectra arising at the two- and three-phonon thresholds. (From Ref. [27]. © 2003 by the American Physical Society)